\documentclass[aip,apl,amsmath,
twocolumn,
]{revtex4}
\usepackage[english]{babel}
\usepackage{graphicx}
\usepackage{dcolumn}
\usepackage{amsfonts}
\usepackage{amsmath}
\usepackage{bm}
\usepackage[colorlinks=true, pdfstartview=FitV, linkcolor=blue, citecolor=blue, urlcolor=blue]{hyperref}

\begin{document}

\title{Feedback cooling of cantilever motion using a quantum point contact transducer}

\author{M. Montinaro}

\affiliation{Department of Physics, University of Basel, Klingelbergstrasse 82, 4056 Basel, Switzerland}
  
\author{A. Mehlin}

\affiliation{Department of Physics, University of Basel, Klingelbergstrasse 82, 4056 Basel, Switzerland}
  
\author{H. S. Solanki}

\affiliation{Department of Physics, University of Basel, Klingelbergstrasse 82, 4056 Basel, Switzerland}
  
\author{P. Peddibhotla}

\affiliation{Department of Physics, University of Basel, Klingelbergstrasse 82, 4056 Basel, Switzerland}
  
\author{S. Mack}

\affiliation{Center for Spintronics and Quantum Computation, University of California, Santa Barbara, California 93106, USA}

\author{D. D. Awschalom}

\affiliation{Center for Spintronics and Quantum Computation, University of California, Santa Barbara, California 93106, USA}

\author{M. Poggio}

\affiliation{Department of Physics, University of Basel, Klingelbergstrasse 82, 4056 Basel, Switzerland}

\date{\today}


\begin{abstract} 
  We use a quantum point contact (QPC) as a displacement transducer to
  measure and control the low-temperature thermal motion of a nearby
  micromechanical cantilever. The QPC is included in an active
  feedback loop designed to cool the cantilever's fundamental
  mechanical mode, achieving a squashing of the QPC noise at high
  gain. The minimum achieved effective mode temperature of 0.2 K and
  the displacement resolution of $10^{-11}$ m/$\sqrt{\text{Hz}}$ are
  limited by the performance of the QPC as a one-dimensional conductor
  and by the cantilever-QPC capacitive coupling.
\end{abstract}



\maketitle


Sensitive displacement transducers are a key component in a wide
variety of today's most sensitive experiments, including precision
measurements of force \cite{Rugar:2004}, mass \cite{Chaste:2012},
gra\-vi\-ta\-tional waves \cite{Whitcomb:2008}, as well as tests of
the macroscopic manifestation of quantum mechanics itself
\cite{Bose:1999}. Sensitive techniques coupling mechanical motion to
optical, microwave, capacitive, magnetic, or piezoelectric effects,
each have advantages in particular applications \cite{Poot:2012}.  The
displacement imprecision of some of these measurements approaches the
standard quantum limit on position detection \cite{Caves:1980}, i.e.\
the limit set by quantum mechanics to the precision of continuously
measuring position. Such exquisite resolution has enabled recent
experiments measuring quantum states of mechanical motion in a
resonator \cite{OConnell:2010, Teufel:2011, Chan:2011}.

It naturally follows that with such fine measurement resolution comes
equally fine control of the mechanical motion, enabling both tuning of
a resonator's linear dynamic range \cite{Kozinsky:2006} and
manipulation of its time response \cite{Poggio:2007}. In fact, such
conditions allow for the application of active feedback cooling
\cite{Poggio:2007} as a method for preparing a mechanical oscillator
near its quantum ground state. Unlike side-band cooling, which has
recently been used to cool high-frequency resonators into their ground
state \cite{Teufel:2011, Chan:2011}, feedback cooling is particularly
well-suited to the ultra-soft low-frequency cantilevers typically used
in sensitive force measurements. The minimum phonon occupation number
achieved by this method depends only on the detector's displacement
imprecision and the resonator's thermal noise \cite{Poggio:2007}. As a
result, a widely applicable transduction scheme with low displacement
imprecision has the potential to prepare resonators in quantum states
of mechanical motion.

Here we investigate one such technique: the use of a quantum point
contact (QPC) as a sensitive detector of cantilever displacement
\cite{Poggio:2008}. The QPC transducer works by virtue of the strong
dependence of its conductance on disturbances of the nearby electric
field by an object's motion. In particular, a QPC is advantageous due
to its versatility as an off-board detector, its applicability to
nanoscale oscillators, and its potential to achieve quantum-limited
detection \cite{Clerk:2003, Clerk:2004}. Most other displacement
detection schemes require the functionalization of mechanical
resonators with electrodes, magnets, or mirrors \cite{Poot:2012}. These requirements
tend to compete with the small resonator mass and high quality factor
necessary to achieve low thermal noise and high coupling strength to
the detector. Since all resonators disturb the nearby electric field,
the QPC transducer, in principle, requires no particular
functionalization. The coupling of a mechanical resonator to a QPC
device is also interesting as one of a series of new hybrid systems
coupling mechanical resonators with microscopic quantum systems. In
particular, such a system may be the first step towards coupling a
resonator with an off-board quantum dot, in an approach aimed at the quantum
control of mechanical objects, precision sensing, and quantum
information processing.

\begin{figure}[tp]
	\includegraphics[width=8cm]{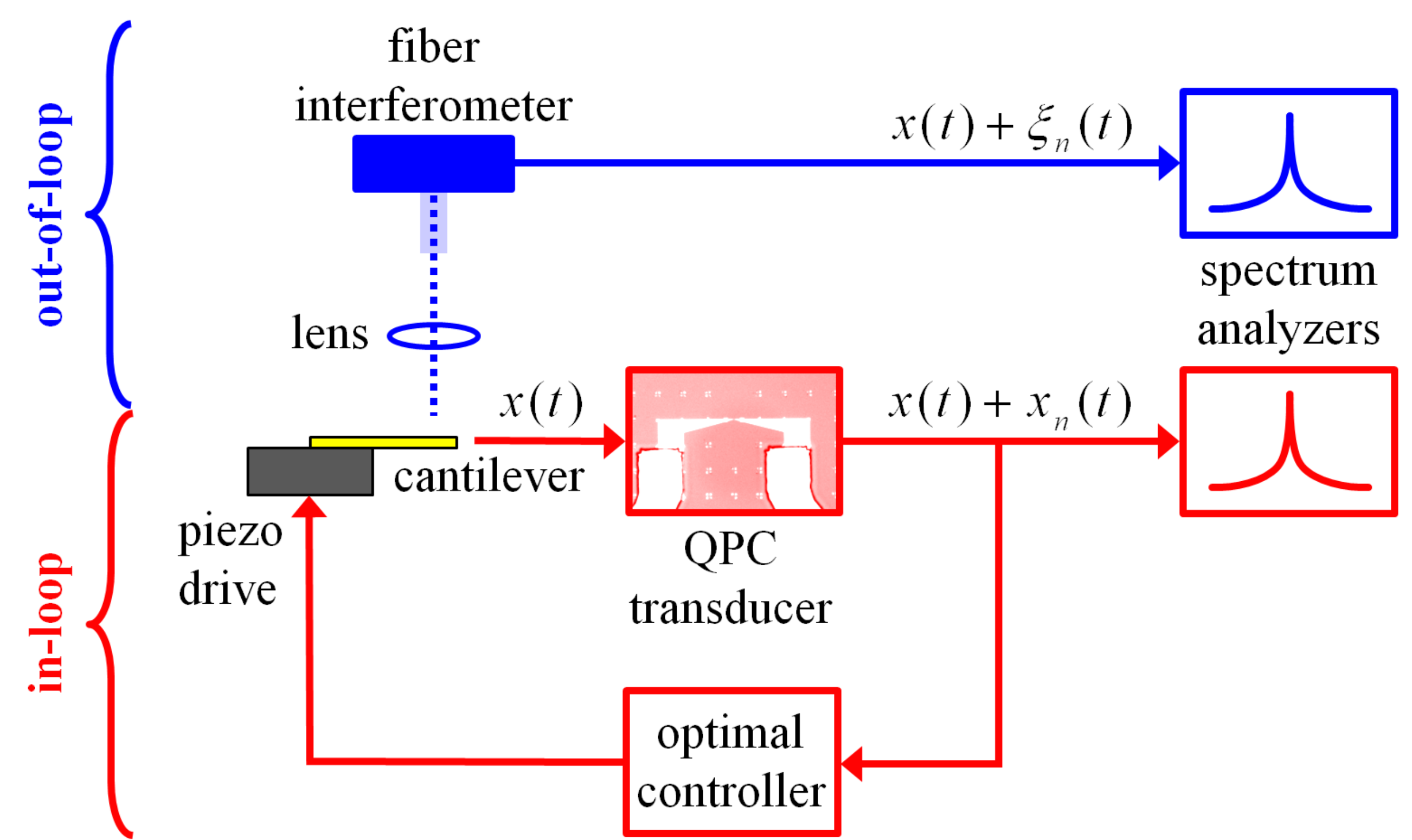}
	\caption{\label{f:fb} Schematic diagram of the experimental setup. In the red loop, the motion of the
	cantilever is transduced by a quantum point contact and amplified by an optimal controller, before
	being sent to a piezoelectric element mechanically coupled to the cantilever. The motion is also
	independently detected by an out-of-loop fiber interferometer, shown in blue.}
\end{figure}

The ex\-pe\-ri\-men\-tal setup described in this work is shown
schematically in Fig.~\ref{f:fb}: the QPC transducer generates an
electrical signal proportional to the cantilever displacement; such a
signal is then amplified by a digital optimal controller
\cite{Optcontrol} and sent to a piezoelectric element mechanically
coupled to the cantilever.  We choose the phase of the optimal control
feedback such that the cantilever oscillation is damped. Here we
demonstrate the possibility to damp the thermal noise spectrum of the
resonator below the QPC measurement noise floor, which is close to the
shot noise level. Such an effect has already been demonstrated for an
opto-electronic loop \cite{Buchler:1999, Bushev:2006, Poggio:2007,
  Lee:2010} and is known as intensity noise ``squashing''. In such a
regime, the effect on the motion of the resonator can be further
validated by detecting it outside the feedback loop, by a second
transducer whose measurement noise is not correlated with the motion.
In this work, such an out-of-loop measurement has been carried out by
means of a low-power laser interferometer.

The QPC transducer is made from a heterostructure grown by
molecular-beam epi\-taxy on a (001) GaAs substrate; the structure consists of
a 600~nm GaAs layer grown on top of the substrate, followed by 20~nm
Al$_{0.25}$Ga$_{0.75}$As, a Si delta-doped layer, 40~nm
Al$_{0.25}$Ga$_{0.75}$As and finally a 5~nm GaAs cap. The 2DEG lies
only 65~nm below the surface and is characterized by a carrier density
$n = 2.5 \times 10^{11}\,\mbox{cm}^{-2}$ and mobility $\mu =
10^5\,\mbox{cm}^2\mbox{V}^{-1}\mbox{s}^{-1}$ at $T = 4.2$~K.  Ti/Au
(5/15~nm) split gates patterned by electron-beam lithography define
the QPC within the 2DEG. The application of a negative
potential $V_g$ between the gates and the 2DEG forms a variable-width
channel through which electrons flow. Ni/Ge/Au/Ni (2/26/54/15~nm)
ohmic contacts are defined on either side of the channel, across which
an applied source-drain voltage $V_{sd}$ drives the QPC conductance.

The micromechanical resonator is a commercial cantilever (Arrow TL1
from NanoWorld AG) made from monolithic silicon which is highly doped
to make it conductive. The cantilever consists of a $(500 \times 100
\times 1)\,\mu$m shaft ending with a triangular sharp tip (radius of
curvature around 10~nm) which has been metallized with Ti/Au
(10/30~nm) to reduce the non-contact friction produced by the
interaction with the QPC sample surface \cite{Stipe:2001}. Due to the
cantilever conductivity, a voltage $V_l$ can be applied to its tip by
contacting the base of the cantilever chip. At $T = 4.2$~K, the
cantilever has a resonant frequency $\nu_0 = 7.9$~kHz and an intrinsic
quality factor $Q_0 = 2.0 \times 10^5$, measured using a ``ring-down''
technique, by exciting the cantilever and measuring the decay of its
oscillation amplitude. The oscillator spring constant is determined to
be $k = 2\times10^{-3}$~N/m through measurements of its thermal noise
spectrum at several different temperatures.

The cantilever and QPC are mounted in a vacuum chamber with a pressure
below $10^{-6}$~torr at the bottom of a $^4$He cryostat ($T = 4.2$~K),
which is isolated from environmental vibrations. A two-tesla magnetic field,
perpendicular to the QPC surface, is applied in order to reduce the
backscattering of electrons in the conductance channel, thus providing
a steeper conductance quantization; the field also has the effect of
further damping the external vibrations of the system. A
three-dimensional positioning stage with nanometer precision and
stability (Attocube AG) moves the QPC relative to the cantilever.

\begin{figure}[tp]
	\includegraphics[width=8.5 cm]{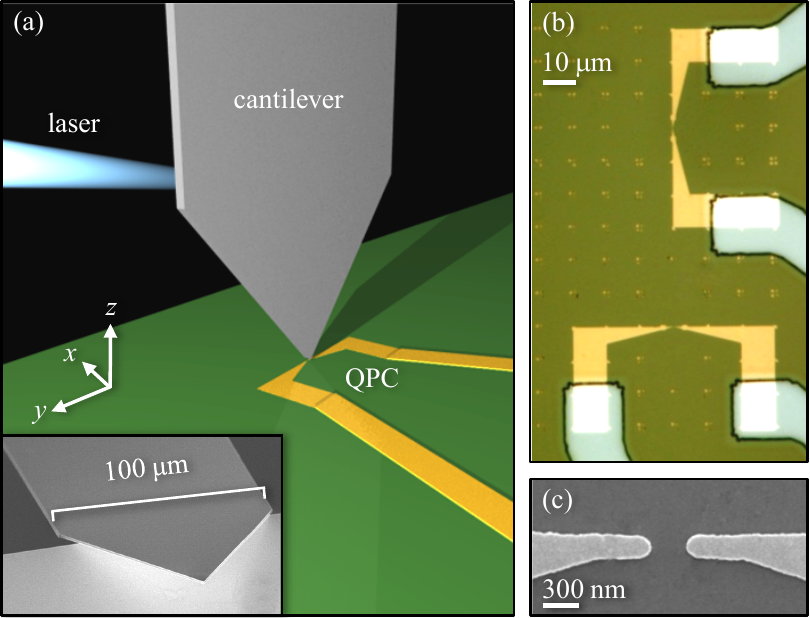}
	\caption{\label{f:setup} (a) Schematic picture of the
          experimental setup. The inset is a scanning electron
          micrograph of the cantilever tip, mounted on a silicon
          nitride blade to damp its vibrations. (b) Optical micrograph
          of the device containing two QPCs in different orientations
          with respect to the oscillation direction of the
          cantilever. (c) Scanning electron micrograph of the active
          region of the QPC.}
\end{figure}

The displacement measurement is made by positioning the tip of the
cantilever about 80~nm above the QPC, as shown schematically in
Fig.~\ref{f:setup}a. Owing to the proximity of the cantilever to the
QPC itself, the cantilever's tip and the QPC are capacitively
coupled. The tip acts as a movable third gate above the device
surface, able to affect the potential landscape of the QPC channel and
thereby to alter its conductance $G$. A voltage $V_g$ applied to the
two gates patterned on the surface modifies $G$ in the same manner.

\begin{figure}[tp]
	\includegraphics[width=6.8 cm]{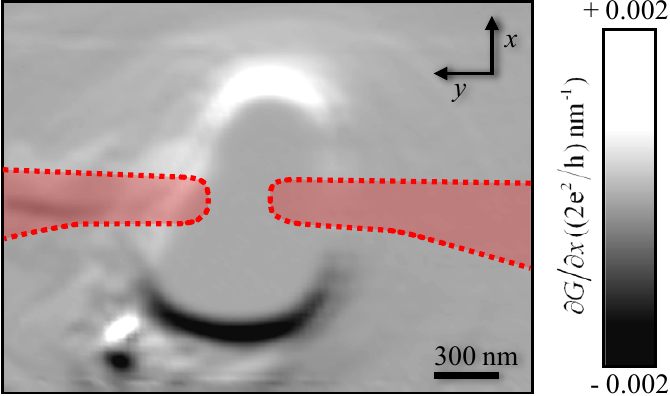}
	\caption{\label{f:Gmap} $\partial G/\partial x$ plotted as a
          function of the cantilever position over the QPC device, at
          a fixed distance $z = 80$~nm. The red shaded areas show the
          position of the QPC gates. $V_g = -0.837$~V, $V_l =
          -1.280$~V. Data were processed with the WSxM software
          \cite{Horcas:2007}.}
\end{figure}

The tip-QPC capacitive coupling strongly depends on their relative separation.
Furthermore, the sensitivity of $G$ to the cantilever motion also
depends on the relative orientation between the direction along which
the cantilever oscillates and the one followed by the current
flow. For studying this behavior, different QPCs have been defined on
the same chip (Fig.~\ref{f:setup}b), with the split gates patterned
such that a current flows either along the cantilever's oscillation
direction, or perpendicular to it; we have found the former to be the
best orientation. In order to map the effect on the conductance by the
position of the cantilever above the QPC device, $G$ has been recorded
while scanning the cantilever at fixed distance $z$, with a potential
$V_l$ applied. In such a conductance map, the position corresponding
to the highest sensitivity is where the absolute value of the spatial
derivative along the oscillation direction is maximum, as shown in
Fig.~\ref{f:Gmap}.

\begin{figure}[htbp]
	\includegraphics[width=8.5 cm]{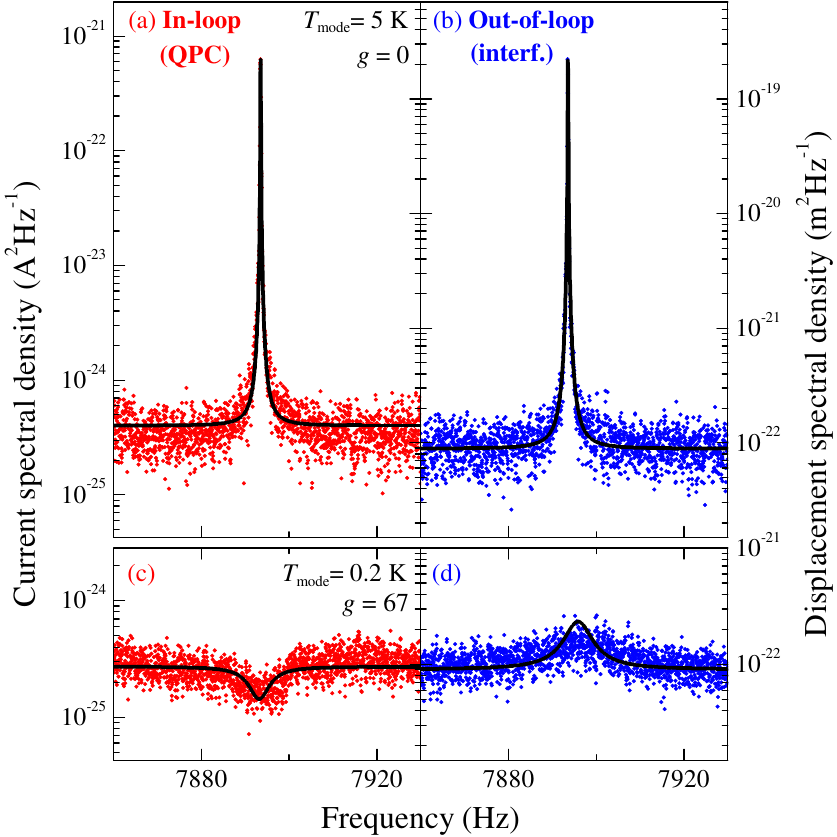}
	\caption{\label{f:spectra} Cantilever fundamental mode spectrum detected at a
	base temperature $T = 4.2$~K by (a, c) a QPC transducer, (b, d) a fiber
	interferometer. The QPC response is expressed
	in terms of both A$^2$~Hz$^{-1}$ (left axis) and m$^2$~Hz$^{-1}$ (right axis).
	(a) and (b) represent the cantilever thermal noise, (c) and (d) are the results
	of the application to the cantilever of a damping feedback force. In the measurements
	shown here, the QPC gives a conductance response of $2\times10^{-4}\,(2e^2/h)\,\mbox{nm}^{-1}$
	of cantilever motion.}
\end{figure}

With the tip of the cantilever so positioned, the QPC acts as a
transducer of the cantilever thermal motion.  Its displacement
resolution, without any feedback force, is shown in
Fig.~\ref{f:spectra}a, along with that of the low-power laser
interferometer used in the out-of-loop measurement, shown in
Fig.~\ref{f:spectra}b.  The resonances represent the cantilever
fundamental mode and match in both frequency and quality factor. A DC
source-drain voltage $V_{sd} = 5.0$~mV drives a current through the
QPC with voltages $V_g = -0.837$~V applied to the gates and $V_l =
-1.280$~V to the cantilever. This configuration defines a conductance
corresponding to one half the value of the first conductance quantum
$2e^2/h$, where $e$ and $h$ are the electron charge and Plank's
constant respectively. Under the same conditions, we also measure the
cantilever displacement using an optical fiber interferometer.  The
interferometer consists of 20~nW of laser light from a
temperature-tuned 1,550-nm distributed feedback laser diode focused
onto a region close to the cantilever tip and then reflected back onto
the cleaved end of an optical fiber \cite{Bruland:1999}.  The fiber
end is coated with 25~nm of Si for optimal reflectivity.  In order to
express the QPC current response (left axis in Fig.~\ref{f:spectra}a)
in terms of cantilever motion (right axis), we have normalized the
peak QPC current spectral density to the peak of the displacement
response measured by the interferometer, obtaining a conductance
response up to $0.002\,(2e^2/h)\,\mbox{nm}^{-1}$ of cantilever motion.

For frequencies in the vicinity of the fundamental resonance mode, the
motion of a cantilever is well approximated by the equation of a
damped harmonic oscillator, driven by thermal force and, in case of a
closed-loop system, also by a feedback force.  In this work, we
approximate the optimal control operated in the feedback loop as a
force proportional to the displacement with a $\pi / 2$ phase lag. In
the experiment, the phase of the feedback signal is affected by the
delay introduced by stray capacitances in the loop and it has been
tuned in order to achieve the desired value $\pi / 2$ for optimal
damping \cite{Poot:2012}. The equation of motion of the cantilever can
thus be written as:
\begin{equation}
  m\ddot{x} + \Gamma_0\dot{x} + kx = F_{\text{th}} - g \Gamma_0
  \omega_0\,\delta(t-\pi/(2 \omega_0))\otimes(x+x_n),
	\label{e:motion}
\end{equation}
where $x(t)$ is the displacement of the oscillator as a function of
time, $m$ is the oscillator effective mass, $\omega_0$ is its angular
resonance frequency, $\Gamma_0 = m \omega_0/Q_0$ is its intrinsic
dissipation, $k = m\omega_0^2$ is its spring constant, $F_{\text{th}}$
is the random thermal Langevin force, $g$ is the feedback gain
coefficient, $x_n(t)$ is the measurement noise on the displacement
signal, $\delta$ is the Dirac distribution, and the symbol $\otimes$
denotes convolution.

Considering in (\ref{e:motion}) frequency components of the form
$\hat{F}_{\text{th}}(\omega)\,e^{i\omega t}$ and
$\hat{x}_n(\omega)\,e^{i\omega t}$, it is possible to determine the
resonator displacement spectral density as measured in-loop
($S_x^{\text{il}}$) or out-of-loop ($S_x^{\text{ol}}$). To do so, we
have followed the procedure described in Refs. \cite{Poggio:2007,
  Poot:2012}: the former involves the calculation of the white
spectral density of the thermal force $F_{\text{th}}$ through the
application of the fluctuation-dissipation theorem.  The out-of-loop
response is simply the sum of the actual displacement of the
cantilever $S_x$ and the white spectral density of the interferometer
measurement noise $S_{\xi_n}$.  On the other hand, in the case of the
in-loop response, feedback produces anticorrelations between the
transduction noise and the mechanical motion of the cantilever.  The
resulting equations are:
\begin{eqnarray}
	S_x^{\text{il}} & = & \frac{\frac{4\omega_0^3 k_B T}{Q_0 k} 
	+ \left[\left(\omega_0^2 - \omega^2\right)^2 + \left(\frac{\omega_0 \omega}{Q_0}\right)^2\right] S_{x_n}}
	{\left(\omega_0^2 - \omega^2\right)^2 + \left[\frac{\omega_0}{Q_0} \left(\omega + g \omega_0\right)\right]^2}, \label{e:il} \\
	S_x^{\text{ol}} & = & S_x + S_{\xi_n}
	= \frac{\frac{4\omega_0^3 k_B T}{Q_0 k} + \left(\frac{g \omega_0^2}{Q_0}\right)^2 S_{x_n}}
	{\left(\omega_0^2 - \omega^2\right)^2 + \left[\frac{\omega_0}{Q_0} \left(\omega + g \omega_0\right)\right]^2}
	+ S_{\xi_n}, \label{e:ol}
\end{eqnarray}
where $S_{x_n}$ is the white spectral density of the QPC measurement noise $x_n$, $k_B$ the Boltzmann
constant and $T$ the bath temperature.

We fit the undamped in-loop and out-of-loop spectra in
Figs.~\ref{f:spectra}a, b with feedback gain $g = 0$. We first fit the
out-of-loop spectrum using (\ref{e:ol}) with three free parameters:
$\omega_0$, $Q_0$ and $S_{\xi_n}$. Setting these parameters as
constants, we then fit the in-loop spectrum with $S_{x_n}$ as the only
free parameter. Both spectra are well-described by the fit functions.
The value of $Q_0$ extracted from this procedure is equal to $8.0
\times 10^4$ and is lower than that measured with the cantilever far
from the QPC surface, due to unavoidable non-contact
friction. $S_{x_n}$ and $S_{\xi_n}$ express the level of the noise
floors for the in-loop and the out-of-loop measurements, respectively.
They set the resolution of the QPC and the laser interferometer as
displacement transducers, which is roughly the same for both: below
$10^{-11}\,\mbox{m}/\sqrt{\mbox{Hz}}$.

The effective temperature of the fundamental mode does not depend on
the measurement noise and is defined, according to the equipartition
theorem, as:
\begin{equation} \label{e:T}
	T_{\text{mode}} = \frac{k}{2\pi k_B}\,\int^{\infty}_{0}S_x\,d\omega.
\end{equation}
For the data in Figs.~\ref{f:spectra}a and b, the value of
$T_{\text{mode}}$ resulting from the equation above, using the
expression of
$S_x$ obtained from the fit, is equal to 5~K, which
corresponds to the bath temperature $T$ of liquid helium.

We now describe the feedback cooling of the cantilever fundamental
mode using the QPC transducer. Optimal control of the resonator motion
in the feedback loop allows the damping of its fundamental mode
oscillations and therefore the reduction of $T_{\text{mode}}$.  Such
an effect can be described with the application of a non-zero gain $g$
in the equation of motion (\ref{e:motion}). Increasing the value of
$g$ produces anticorrelations between the in-loop transduction noise
and the mechanical oscillator motion \cite{Bushev:2006, Poggio:2007,
  Lee:2010}.  As a consequence, the displacement spectral density
detected inside the feedback loop can even exhibit a dip below its
noise floor near the oscillator's resonant frequency, as shown in
Fig.~\ref{f:spectra}c. This spectrum represents noise ``squashing''
for a transduction scheme limited by electron, rather than photon,
shot-noise. The solid line plotted along with this in-loop spectrum in
Fig.~\ref{f:spectra}c represents a fit computed using (\ref{e:il}),
with the value of $Q_0$ extracted previously and with $g$ as the free
parameter.

In order to provide a validation of the observed phenomenon and an
independent measurement of $T_{\text{mode}}$, the cantilever motion is
also detected through the out-of-loop laser interferometer.  This
spectrum, shown in Fig.~\ref{f:spectra}d, exhibits a peak above the
uncorrelated measurement noise $S_{\xi_n}$. In order to compare our
model with the measured data, we plot (\ref{e:ol}) as a solid line in
Fig.~\ref{f:spectra}d, using $Q_0$, $S_{x_n}$, and $S_{\xi_n}$
extracted from previous fits and $g$ extracted from the fit to the
damped in-loop QPC spectrum of Fig.~\ref{f:spectra}c.  The plot of the
out-of-loop spectrum highlights the agreement between our theoretical
model and the experimental data.

To calculate the mode temperature, a general expression can be
elaborated from (\ref{e:T}), using the expression given in
(\ref{e:ol}) for $S_x$; we find for $T_{\text{mode}}$ the same result
obtained in Ref.~\cite{Poggio:2007}, valid for a high quality factor:
\begin{equation} \label{e:T2}
	T_{\text{mode}} = \frac{T}{1 + g} + \frac{k \omega_0}{4k_B Q_0} \left(\frac{g^2}{1 + g}\right) S_{x_n}.
\end{equation}
The values of $T_{\text{mode}}$ resulting either from direct
integration of the spectrum as in (\ref{e:T}), or by extracting the
parameters from the fit and then substituting them into (\ref{e:T2}),
are equal within our precision: 0.2~K, twenty times less than the
bath temperature.

Equation (\ref{e:T2}) implies that, in the limit $g \gg 1$, the minimum
achievable temperature is:
\begin{equation} \label{e:Tmin}
	T_{\text{mode}}^{\text{min}} = \sqrt{\frac{m \omega_0^3 T}{k_B Q_0} S_{x_n}},
\end{equation}
which in our case results to be $(0.21 \pm 0.05)$~K, equal whithin the
error to the observed value of $T_{\text{mode}}$ in the noise
squashing regime.

In order to achieve the lowest possible mode temperature, thus
accessing a state with a low occupation number ($N_{\text{mode}} = k_B
T_{\text{mode}} / \hbar \omega_0$), future experiments should employ
cantilevers with a low mass, low resonance frequency, and a high
quality factor. The base temperature should also be lowered, by means
of a $^3$He or a dilution refrigerator. In addition, lowering the
measurement noise floor $S_{x_n}$ would represent a crucial
improvement, involving both a decrease in the QPC current noise and an
increase in the sensitivity of the QPC to the cantilever displacement.
In the experiment presented here, the QPC noise floor is within a
factor 10 above its current shot-noise limit; an improvement of the
measurement setup would allow us to approach this limit. On the other
hand, a better sensitivity could be achieved in two ways: improving
the performance of the QPC as a one-dimensional conductor, and
increasing the cantilever-QPC capacitive coupling. The former implies
using a QPC defined on a 2DEG with a higher electron mobility and at a
lower bath temperature.  The effect would be to obtain a sharper QPC
conductance quantization, and therefore a higher sensitivity to local
electrostatic fields. The latter requires bringing the conductance
channel closer to the cantilever tip, by using a QPC defined on a
shallower 2DEG, optimizing the shape of the tip for a higher influence
on the QPC potential landscape, and reducing shielding effects from
both charged defects and gates on the QPC surface.  

We acknowledge support from the Canton Aargau, the Swiss NSF (grant
200020 140478), the NCCR QSIT, and the US NSF.


\end{document}